\documentclass[final,3p,times,twocolumn]{elsarticle}
\usepackage{ulem}
\usepackage{amssymb}
\usepackage{amsmath}
\usepackage{subcaption}
\usepackage{graphicx}
\usepackage{xcolor}

\usepackage{hyperref}
\journal{Ultrasonics}
\DeclareMathOperator{\RPSF}{RPSF}

\begin{document}

\begin{frontmatter}

%% Title, authors and addresses

%% use the tnoteref command within \title for footnotes;
%% use the tnotetext command for theassociated footnote;
%% use the fnref command within \author or \affiliation for footnotes;
%% use the fntext command for theassociated footnote;
%% use the corref command within \author for corresponding author footnotes;
%% use the cortext command for theassociated footnote;
%% use the ead command for the email address,
%% and the form \ead[url] for the home page:
%% \title{Title\tnoteref{label1}}
%% \tnotetext[label1]{}
%% \author{Name\corref{cor1}\fnref{label2}}
%% \ead{email address}
%% \ead[url]{home page}
%% \fntext[label2]{}
%% \cortext[cor1]{}
%% \affiliation{organization={},
%%             addressline={},
%%             city={},
%%             postcode={},
%%             state={},
%%             country={}}
%% \fntext[label3]{}

\title{Evaluation of effective wave velocities 
in polycrystalline materials using the ultrasonic reflection matrix} %% Article title

%% use optional labels to link authors explicitly to addresses:
%% \author[label1,label2]{}
%% \affiliation[label1]{organization={},
%%             addressline={},
%%             city={},
%%             postcode={},
%%             state={},
%%             country={}}
%%
%% \affiliation[label2]{organization={},
%%             addressline={},
%%             city={},
%%             postcode={},
%%             state={},
%%             country={}}

\author[1,2]{Gatien CLEMENT}
%\ead{gatien.clement@espci.fr}
\author[1]{Alexandre AUBRY}
\author[2]{Cécile BRÜTT}
\author[2]{Benoît GÉRARDIN}
\author[1]{Claire PRADA\corref{cor1}}
\ead{claire.prada@espci.fr}

\cortext[cor1]{Corresponding author}

\affiliation[1]{organization={Institut Langevin, ESPCI Paris, PSL University, CNRS},%Department and Organization
            %addressline={1 Rue Jussieu},
            postcode={75005},  
            city={Paris},
            country={France}}
\affiliation[2]{organization={Safran Tech, Digital Sciences \& Technologies Department},%Department and Organization
            addressline={Rue des Jeunes Bois, Châteaufort},
            postcode={78114},  
            city={Magny-les-Hameaux},
            country={France}}

%% Abstract
\begin{abstract}
%% Text of abstract
%The volumetric (volumetric a trait à la mesure du volume ..... je n'aime pas trop)
In-depth characterization of heterogeneous materials has long been a challenge in non-destructive testing. Here, a method is proposed to determine the elastic constants of metallic polycrystalline materials using back-scattered ultrasound. The waves scattered by the microstructure are analyzed to image the effective bulk velocities. To this end, a reflection matrix is acquired with an array of transducers. The projection of this matrix onto a focused basis is used to  estimate an average point spread function.  Optimizing this function with respect to the propagation model leads to an estimation of the longitudinal velocity. Additional treatments are developed to adapt the method to map the shear wave velocity. The local Poisson's ratio is then deduced from the ratio between those two velocities. Young's modulus and shear modulus can also be obtained assuming known densities. This matrix approach is experimentally validated on different polycrystalline materials. A sample displaying heterogeneous mechanical properties is then simulated to assess the accuracy and the resolution of the method. Its strengths and limitations are discussed, demonstrating its potential for quantitative non-destructive material characterization.
\end{abstract}

%%Graphical abstract
%\begin{graphicalabstract}
%\includegraphics{grabs}
%\end{graphicalabstract}

%%Research highlights
%\begin{highlights}
%\item Research highlight 1
%\item Research highlight 2
%\end{highlights}

%% Keywords
\begin{keyword}
%% keywords here, in the form: keyword \sep keyword
Ultrasonic reflection matrix, Polycrystalline materials, Wave velocity estimation, Elastic constants, Non-destructive testing
%% PACS codes here, in the form: \PACS code \sep code

%% MSC codes here, in the form: \MSC code \sep code
%% or \MSC[2008] code \sep code (2000 is the default)

\end{keyword}

\end{frontmatter}

%% Add \usepackage{lineno} before \begin{document} and uncomment 
%% following line to enable line numbers
%% \linenumbers

%% main text
%%

%% Use \section commands to start a section
\section{Introduction}
\label{sec_intro}

%% Labels are used to cross-reference an item using \ref command.

In the field of non-destructive testing (NDT), ultrasound imaging is widely employed for flaw detection and material characterization. In particular, the computation of a reflectivity image from back-scattered echoes,  can provide precise localization of defects. However, the accuracy of such imaging techniques is critically dependent on prior knowledge of the speed of wave  within the medium. In polycrystalline materials, wave propagation can be highly complex, with scattered echoes generally  treated as undesirable noise. Often referred to as speckle in medical ultrasound, these scattered signals can actually be used as a source of information. Here, we focus on material characterization with quantitative ultrasound imaging. We propose a method taking  advantage of the waves back-scattered by the microstructure to image the effective bulk velocities and deduce the elastic properties of metal alloys.

The velocities of elastic waves in a material is a widely used indicator of its mechanical properties. Resonant ultrasound spectroscopy, for example, is a well-established reference technique that determines elastic constants by analyzing the resonance frequencies of small calibrated samples \cite{leisure_resonant_1997}. Although highly precise, this method is inherently destructive. A more conventional nondestructive alternative is pulse-echo measurement, where wave velocity is determined by analyzing reflections by the sample interfaces \cite{anderson_direct_1998}. This technique requires a good knowledge of the sample geometry and only provides average velocities estimates, thus preventing any local velocity estimation. Mapping elastic properties within a material would be particularly valuable for characterizing inhomogeneous samples or in cases where pulse-echo methods fail due to high attenuation or complex geometries. This approach could facilitate the detection of defects that are difficult to detect in conventional reflection images but exhibit different elastic properties, such as some inclusions \cite{cen_inclusions_2019}, or microstructural changes within the same material  \cite{lan_experimental_2014}.

The estimation of spatially varying wave velocities from surface data has been extensively studied in different fields. In geophysics, various inversion techniques have been developed, often relying on layered media models \cite{aki_quantitative_2009}. Migration techniques are also used to efficiently compute velocity maps and improve image quality \cite{grechka_encyclopedia_2014}, relying on interface models,  the types of heterogeneities that can be resolved by these methods are limited. 
In medical imaging, speed-of-sound mapping has gained significant attention due to its dual role: enhancing image quality by correcting aberrations and serving as a quantitative tool for diagnosing pathologies. Recent advances in this field have followed two main approaches. Data-driven methods that leverage deep learning \cite{feigin_deep_2020,heller_speed--sound_2023} have demonstrated impressive resolution and accuracy, but require extensive training datasets and may suffer from robustness issues due to dataset biases and ground truth limitations. Alternatively, physics-based approaches %\cite{jaeger_computed_2015}- \cite{HeriardDubreuil2026} 
\cite{jaeger_computed_2015, Staehli2020,Ali2022, HeriardDubreuil2023, bureau_self-portrait_2026,Simson2026, HeriardDubreuil2026} 
enable robust velocity mapping with minimal prior assumptions, making them attractive for heterogeneous and weakly known media.

Given the complex geometries and the great variety of materials encountered in NDT, physics-based methods appear to be the best candidates for speed mapping. Consequently, we propose a new approach adapted to solid media from the method previously introduced in medical imaging \cite{bureau_self-portrait_2026}. 
This method exploits the spatial diversity of echoes reflected by small scatterers within the medium and captured by a multi-element array through the analysis of the focused reflection matrix \cite{lambert_reflection_2020}. In metals, the microstructure can generate similar signals. In the single-scattering regime, these signals provide an estimate of the point spread function, which can be optimized to determine the longitudinal wave speed $c_L$. In this paper, developments include the extension of this approach to the estimation of shear wave speed $c_S$. These velocity maps provide the map of the Poisson’s ratio and assuming a constant and known density, the derivation of Young’s and shear moduli maps. 

The feasibility of this method is validated on three different polycrystalline materials using a linear transducer array. Then, its precision and resolution is assessed by a numerical simulation involving a scattering medium with spatially-distributed mechanical properties. The strengths and limitations of the approach are identified, demonstrating its potential as a tool for quantitative non-destructive material characterization.

\section{Measurement of the effective longitudinal wave speed}

\label{sec_longi}

The method, originally developed for medical imaging, takes advantage of both spatial and temporal degrees of freedom provided by a transducer array. Using the focused reflection matrix \cite{lambert_reflection_2020}, these degrees of freedom can be fully exploited to extract meaningful information from the backscattered signals. In particular, Bureau \textit{et al.}~\cite{bureau_self-portrait_2026} demonstrated that the speed of sound can be measured using the focused reflection matrix in biological tissues. In this section, we adapt this approach to metallic materials, where the backscattered signals originate from the interaction with the microstructure, and focus on the measurement of the  effective longitudinal wave velocity.

\subsection{Data acquisitions}

\begin{figure}[ht]
    \centering
    \includegraphics[width=0.99\linewidth]{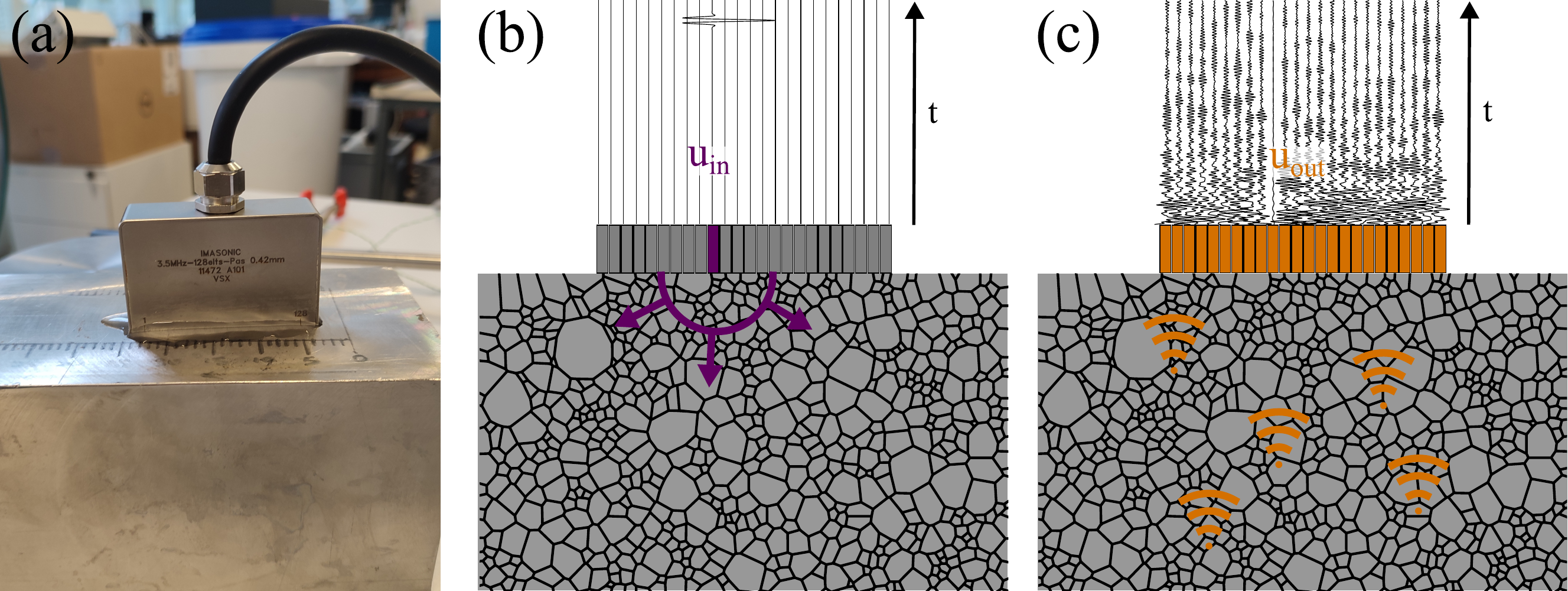}
    \caption{(a) Experimental setup for the acquisition of the reflection matrix on an Inconel 600 sample; principle of the FMC : (b) emission of a pulse with the an element $u_{in}$; (c) reception with all the elements $u_{out}$.}
    \label{fig:setup}
\end{figure}

The experimental setup consists of a linear array of transducers operating at a central frequency of 3.5~MHz and a bandwidth of 3~MHz at -20~dB. The probe is placed in contact with the metallic sample, using water as a coupling medium as shown in Fig.~\ref{fig:setup}.

The reflection matrix is acquired by sequentially emitting a pulse, typically two half-periods at the central frequency, with each individual element, and recording the time-dependent reflected signals $R(\mathbf{u}_{\text{in}}, \mathbf{u}_{\text{out}}, t))$ with all transducers. ${u}_{\text{in}}$ and ${u}_{\text{out}}$ denote the lateral positions of each element acting as source and receiver, respectively. These signals are then arranged into a reflection matrix expressed in the transducer basis:
\begin{equation}
    \mathbf{R}_{uu}(t) = [R({u}_{\text{in}}, {u}_{\text{out}}, t)].
\end{equation}
This acquisition scheme, known as the full matrix capture (FMC) \cite{holmes_post-processing_2005}, is commonly used in non-destructive testing. For the remainder of this paper, the reflection matrix will be analyzed in the frequency domain. Thus, we define
\begin{equation}
    \mathbf{R}_{uu}(f) = \mathcal{F}(\mathbf{R}_{uu}(t)),
\end{equation}
where $\mathcal{F}$ denotes the temporal Fourier transform.

Measurements are conducted on three different metal alloys : Ti64, steel and Inconel 600. The Ti64 sample is a part of a forged billet  with a complex multiscale microstructure with oriented grain flow already studied by Baelde \textit{et al.} \cite{baelde_effect_2018}. No prior information is available on the steel sample. 
The Inconel sample is a  nickel based alloy  that has been studied in Refs.~\cite{shahjahan_comparison_2014,du_burck_attenuation_2026}. It was shown to present a statistically homogeneous and  isotropic distribution of grains of averaged sizes  $90~\mu $m that produce low level of multiple scattering  around 3.5 MHz.

\subsection{The focused reflection matrix}

From the reflection matrix recorded in the transducer basis, the most common approach for imaging is the computation of a confocal image, typically achieved through the Total Focusing Method (TFM). However, TFM only provides a qualitative representation of the medium's reflectivity. Moreover, its reconstruction requires knowledge of the wave speed which is often assumed to be constant. This hypothesis does not hold in heterogeneous materials where local variations (e.g. due to texture~\cite{lan_experimental_2014} or inclusions) can significantly impact wave propagation.

\begin{figure}[ht]
    \centering
    \includegraphics[width=0.6\linewidth]{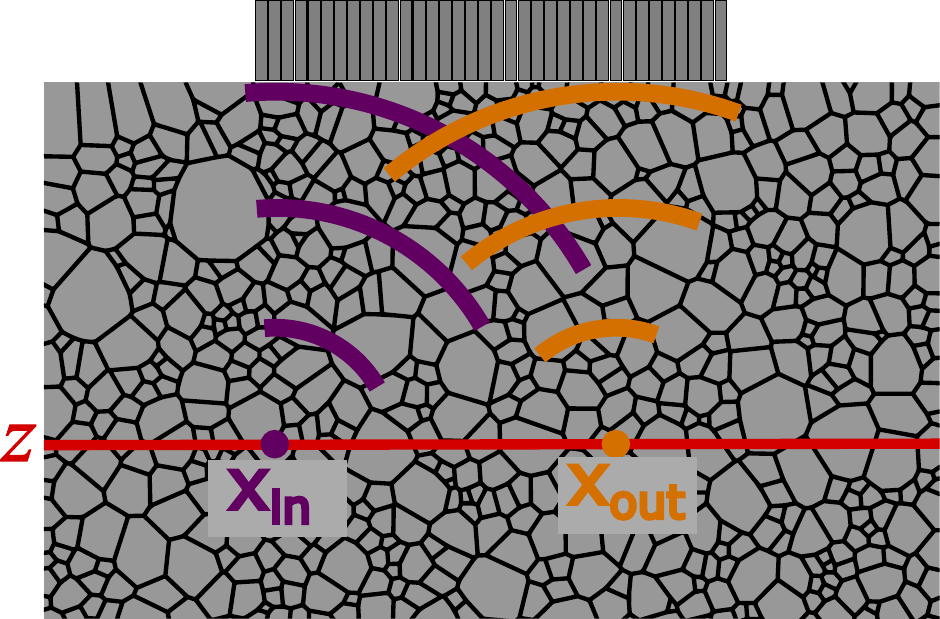}
    \caption{Principle of the reflection matrix in the focused basis. Matrix imaging consists in focused beamforming to probe  points $x_{\text{in}}$ in emission
and $x_{\text{out}}$ in reception at each depth $z$. The set of impulse responses between such virtual transducers form a focused reflection matrix $\mathbf{R}_{xx}(z)$ at each depth $z$.}
    \label{fig:schema_Rrr}
\end{figure}

To cope with the uncertainties on the velocities, the focused reflection matrix~\cite{lambert_reflection_2020} was introduced as part of the ultrasound matrix imaging framework~\cite{lambert_ultrasound_2022}. Unlike confocal imaging, this approach separates the focusing points in transmission and reception, introducing an additional spatial degree of freedom. As a result, the focused reflection matrix provides access to more detailed information about wave propagation by describing the impulse responses between virtual sources and receivers distributed inside the medium, as illustrated in Fig.~\ref{fig:schema_Rrr}.

The transformation of the reflection matrix  from the transducer basis to the focused basis is performed through a projection using the matrix of free-space Green’s function $\mathbf{G}_0(z,f; c)=[G_0(u,x,z,f;c)]$. This matrix encodes the propagation of waves in a  medium of constant wave speed $c$ between the transducer's positions $u$ at medium surface ($z=0$) and the focal points $x$ at depth $z$. Our transducer array is made of rectangular elements of length 10~mm, which is much longer than the average wavelength. A cylindrical acoustic
lens ensures that the emitted and received beams remain collimated in the $(x, z)$ plane. Therefore, 2D Green's functions are considered:

\begin{equation}
    G_0(u,x,z,f;c) = -\frac{i}{4} H_{0} ^{(1)}(k\sqrt{(x-u)^2+z^2}),
    \label{eq:G0}
\end{equation}
where $H_{0} ^{(1)}$ is the Hankel function of the first kind, and $k=2\pi f/c$ is the wavenumber for the assumed wave speed $c$.

Applying this projection at a given depth $z$ for a set of points $x$, the focused reflection matrix $\mathbf{R}_{xx}(z;c)$ is written as in ~\cite{lambert_reflection_2020}:
\begin{equation}
    \mathbf{R}_{xx}(z;c) = \sum_f \mathbf{G}_0^{\dag}(z,f;c) \times  \mathbf{R}_{uu}(f) \times \mathbf{G}_0^{*}(z,f;c).
    \label{eq:beamforming}
\end{equation}
where the symbols $*$ and $\dag$ stands for matrix conjugation and transpose conjugation, respectively.

\begin{figure}
    \centering
    \includegraphics[width=\linewidth]{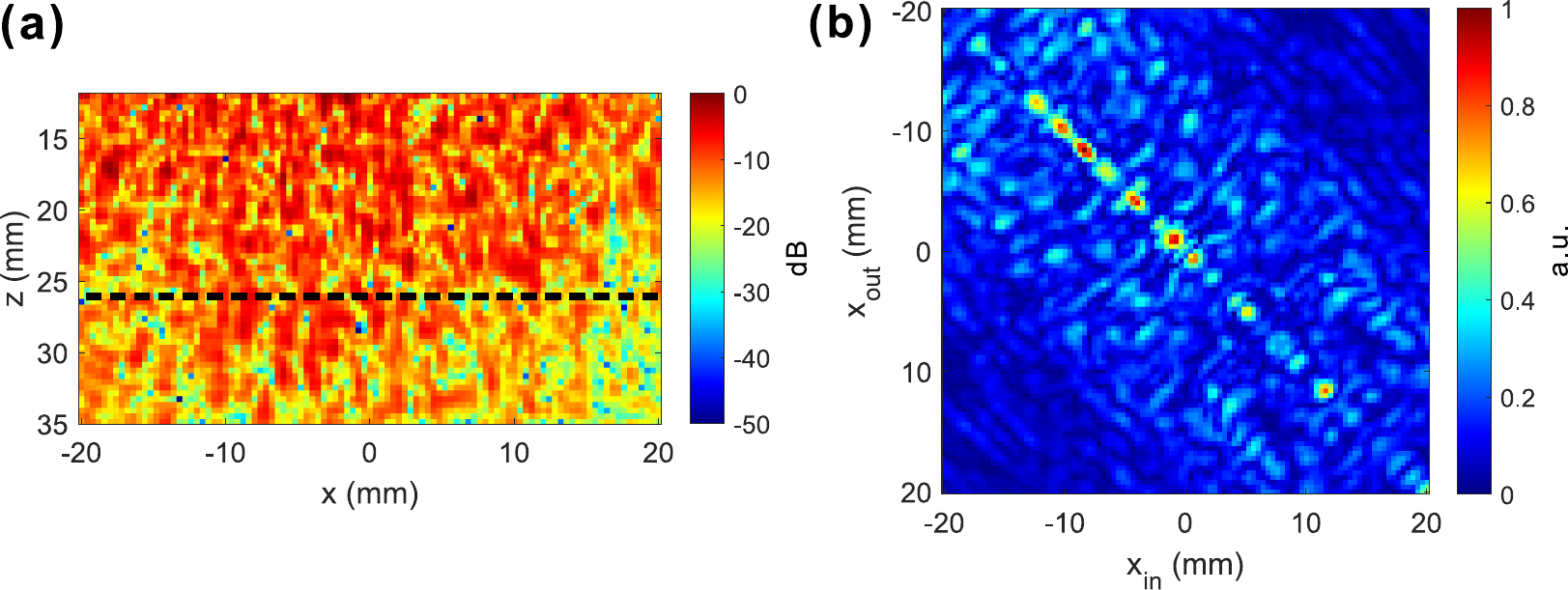}    
\caption{(a) Confocal image in the Inconel 600 sample; (b) corresponding focused reflection matrix at depth $z=26$ mm (purple line in the confocal image).}
\label{fig:Rrr}
\end{figure}

The diagonal of this matrix (${x_{in}}={x_{out}}$) corresponds to the confocal image at depth z. An example of the focused reflection matrix, computed at depth $z=26$~mm in an Inconel 600 sample assuming $c=5840$ m/s, is shown in Fig.~\ref{fig:Rrr}(b), alongside the corresponding confocal image displayed in Fig.~\ref{fig:Rrr}(a). This image reveals a feature that appears random due to backscattering from the microstructure; however, the corresponding wave field is actually deterministic.

Indeed, by analyzing the properties of the focused reflection matrix, valuable information about the microstructure and wave propagation characteristics can be extracted. In the single-scattering regime, when the assumed velocity is correct, most of the energy is concentrated along the main diagonal of the focused reflection matrix. In particular, the spread of this diagonal reflects the width of the focal spot, which depends on diffraction and wave speed mismatch. In order to quantify the off-diagonal energy as a function of the velocity model, we now define the reflection point spread function as previously introduced by Lambert \textit{et al.}~\cite{lambert_reflection_2020}. 
\subsection{The reflection point spread function (RPSF)}
\begin{figure}[h!]
    \centering
    \includegraphics[width=0.99\linewidth]{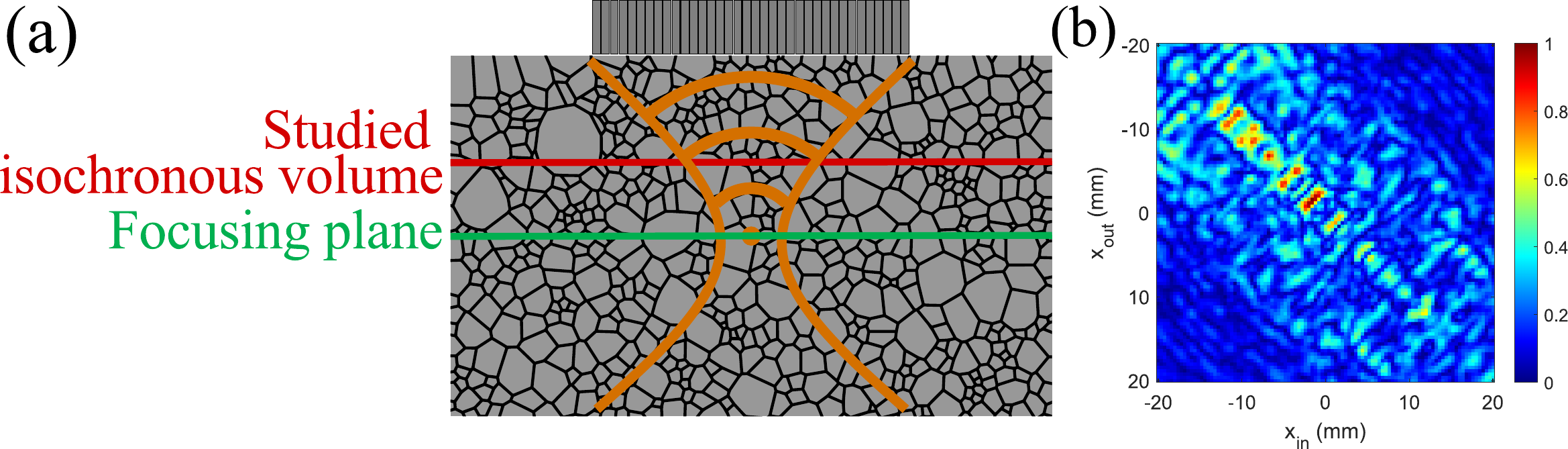}    
    \caption{(a) Illustration of the effect of an incorrect speed hypothesis on the RPSF energy spread; (b) focused reflection matrix obtained for an incorrect velocity assumption of 6080~m/s for the same simulation as in Figure \ref{fig:Rrr}.}
    \label{fig:effect_speed}
\end{figure}

The matrix $\mathbf{R}_{xx}(z)$ implicitly depends on the assumed velocity $c_{0}$ (Eq.~\eqref{eq:G0}). As discussed previously, the width and intensity of signals arising from single scattering events in the microstructure are directly related to the focusing quality. For instance, if the assumed velocity is underestimated, the focal spot is shifted deeper than the targeted depth, leading to a wider spot at the considered depth [Fig.~\ref{fig:effect_speed}(a)].  The impact of a wrong velocity model on the focusing reflection matrix is illustrated in Fig.~\ref{fig:effect_speed}(b). Compared to the reference matrix $\mathbf{R}_{xx}(z)$ beamformed at an adequate wave speed $c$ shown in Fig.~\ref{fig:Rrr}(b), the ratio between the diagonal and off-diagonal echoes is here weaker. This loss of contrast is associated with a wider distribution of the single scattering component around the diagonal of $\mathbf{R}_{xx}(z)$. 

A local analysis of focusing quality can be done by investigating
the reflection point spread function defined as follows:
\begin{equation}
\RPSF(\Delta x,x_{\textrm{in}},t;c)=R(x_{\textrm{in}},x_{\textrm{in}}+\Delta x,z=ct/2; c)
\end{equation}
This function derived from the off-diagonal coefficients of $\mathbf{R}_{xx}$, quantifies
the focusing quality for each speckle grain defined by its spatio-temporal coordinates $ (x_{\textrm{in}}, t)$. The use of the time coordinate instead of the depth coordinate enables to follow the same speckle grain independently of the assumed wave velocity. 

\begin{figure*}[h!]
    \centering
    \includegraphics[width=.8\linewidth]{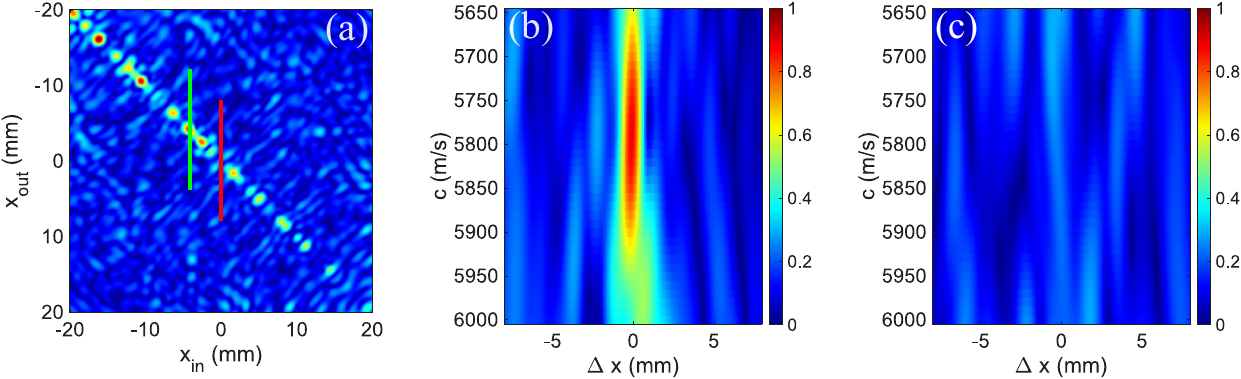}
    \caption{RPSF without spatial averaging in the Inconel 600 sample: (a) focused reflection matrix at $t=10.25$ $\mu s$,  for an hypothesis of $c = 5850$~m.s$^{-1}$; (b) corresponding RPSF at $x=-4$ mm (green line on (a)); (c) corresponding RPSF at $x=0$ mm (red line on (a)).}
    \label{fig:RPSF_raw}
\end{figure*}

Figure~\ref{fig:RPSF_raw} displays the local RPSF associated with two speckle grains at lateral positions $x_{\textrm{in}}$ and time $t=10.25$~$\mu$s and  identified by vertical lines in Fig.~\ref{fig:RPSF_raw}a. While a bright speckle spot can provide a reliable assessment of the focusing quality (Fig.~\ref{fig:RPSF_raw}b) with a clear view on the optimal speed-of-sound (here $\hat{c}=5800$~m.s$^{-1}$), a dark spot leads to a  RPSF modulated by the random distribution of reflectivity. In order to evaluate the quality of the focusing process independently of random reflectivity fluctuations, an  averaging is required to mitigate this effect. Indeed, for isotropic random scattering and under a local isoplanatic assumption, the ensemble
average of the RPSF actually scales as the convolution of the transmit and receive point spread function (PSFs) intensities, $|h_{\textrm{in}}|^2$ and $|h_{\textrm{out}}|^2$, at this location~\cite{bureau_self-portrait_2026}:

\begin{equation}
\left \langle \left | \RPSF(\Delta x,x_{\textrm{in}},t; c ) \right |^2 \right \rangle  \propto  | h_{\textrm{in}} |^2 \stackrel{\Delta x}{\circledast} | h_{\textrm{out}} |^2  (\Delta x ; c).
\end{equation}
where the symbol $\langle . \rangle$ here denotes an ensemble average. Note that, for a FMC acquisition scheme and in virtue of spatial reciprocity, the transmit and receive PSFs are strictly identical: $h_{\textrm{in}}\equiv h_{\textrm{out}}$. An averaged RPSF is therefore a quantitative indicator of the focal spot extension.

To smooth the fluctuations due to the medium random reflectivity, the idea consists in averaging the RPSF over lateral positions and times-of-flight. To maintain a spatial resolution, a local averaging strategy can be followed. The medium is divided into overlapping regions defined by their center $(x_p,t_p)$ and having a spatio-temporal extent provided by a weight function $w(x-x_{p},t-t_p)$  chosen further. A locally averaged reflection point spread function, $\overline{\RPSF}$, is then computed for each region as:
\begin{align}
  &  \overline{\RPSF}(\Delta x,x_p,t_p;c) =  \nonumber \\ & \left\langle |\RPSF ({\Delta x},{x_{in}},t;c)|^2 w({x_{in}}-x_p,t-t_p) \right\rangle_{x_{in},t}^{1/2}.
\end{align}

The size of this averaging region is critical. If it is too small, the averaged RPSF is dominated by speckle fluctuations and remains unreliable. Conversely, an excessively large region reduces spatial resolution. To achieve a suitable compromise between robustness and locality, an adaptive Gaussian weighting is used:
\begin{equation}
    w(x,t)=\exp\left(-\frac{x^2}{2\sigma_x^2}-\frac{t^2}{2\sigma_{t}^2}\right),
\end{equation}
where the lateral and temporal extents $(\sigma_x,\sigma_{t})$ are adjusted according to the local speckle grain size.

\begin{figure}[ht]
    \centering
    \includegraphics[width=0.9\linewidth]{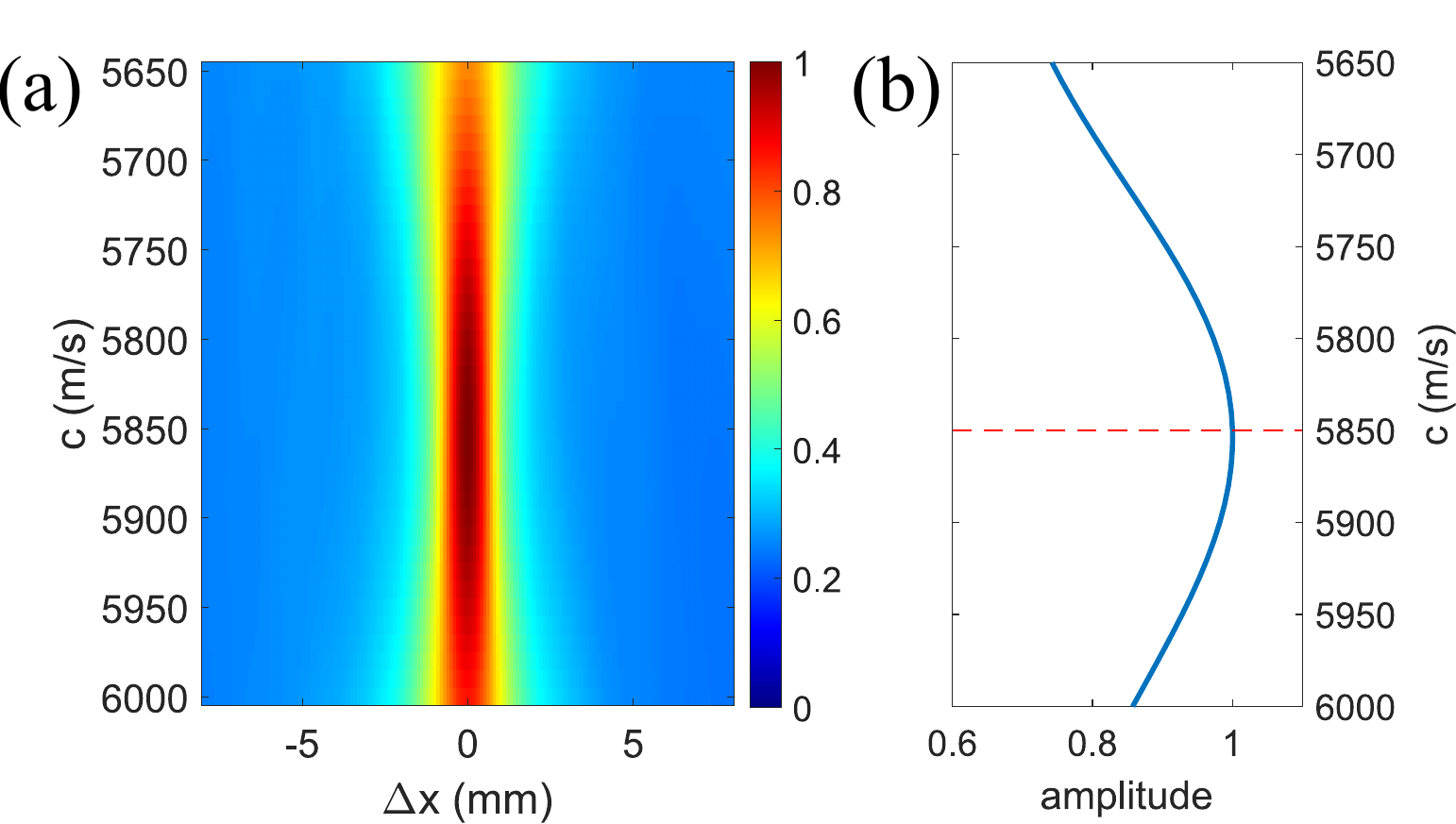}
    \caption{(a) Locally averaged RPSF in inconel 600 at $x=0$~mm and $t=10.25$~$\mu s$ with a Gaussian averaging area such that $\sigma_x = 4l_x = 0.8\lambda = 4.7$~mm and $\sigma_z = 4l_z = 6\lambda = 8.7$~mm; (b) magnitude of the RPSF at $\Delta x = 0$~mm.}
    \label{fig:locally_averaged_RPSF}
\end{figure}

This speckle grain size is estimated from the local autocorrelation of the confocal image. The widths at half-maximum of the autocorrelation peak provide estimates of the resolution cell dimensions $l_x$ and $l_t$, from which:
\[
\sigma_x = N_x\,l_x, \qquad \sigma_{t} = N_{t}\,l_{t},
\]
with $N_x$ and $N_{t}$ controlling the number of independent speckle grains included in the averaging process. 
Figure~\ref{fig:locally_averaged_RPSF} illustrates the result of a locally averaged RPSF with $N_x=N_{t} =4$. Those values were found to be optimal to smooth the RPSF fluctuations and guarantee a sharp estimation of the speed-of-sound at a satisfying resolution.

The optimal velocity at a given position is finally obtained by maximizing the RPSF intensity over a set of velocity hypotheses: 
\begin{equation}
\hat{c}(x_p,t_p)=\underset{\boldsymbol{c}}{\textrm{argmax}}  \left \lbrace  \overline{\RPSF}(\Delta x=0, x_p,t_p;c ) \right \rbrace.
\end{equation}
The time-dependence of this optimal speed-of-sound can be converted into a depth-dependent velocity profile, through the following change of variable: $z_p=\hat{c}(x_p,t_p)t_p/2$.  In the example of Fig.~\ref{fig:locally_averaged_RPSF}, we find $\hat{c}=5850$~m.s$^{-1}$. Repeating this estimation over the entire field of view enables the construction of an optimal velocity map.

\subsection{Longitudinal velocity mapping applied to a statistically homogeneous polycrystalline sample}

\begin{figure}[ht] 
\centering 
\includegraphics[width=0.99\linewidth]{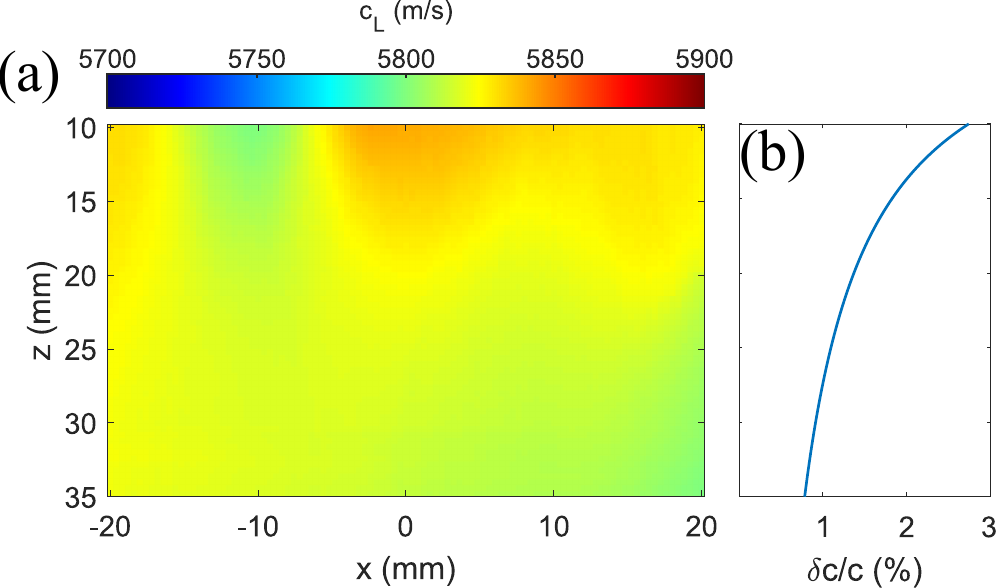} 
\caption{(a) Longitudinal velocity map measured in the Inconel 600 sample; (b) estimated relative uncertainty associated to each depth.
Since the medium is assumed to have homogeneous effective velocity, variations are here interpreted as estimation errors.} 
\label{fig:velocity_map} 
\end{figure}

We apply this approach to an Inconel 600 sample, which is known to be statistically homogeneous and isotropic. Under this assumption, any variation observed in the velocity map is attributed to estimation errors rather than  material inhomogeneities. Figure \ref{fig:velocity_map} displays the  velocity distribution within the sample. The average velocity is found equal to $5820\pm7 $ m/s, close to the velocity measured using a conventional pulse echo technique, $5848 \pm 13$ m/s. This value is also close to the velocities estimated from the head wave measured by laser ultrasonic technique on the same sample and ranging from 5850 m/s at 3 MHz to 5900 m/s at 4 MHz \cite{du_burck_attenuation_2026}.

Additionally, in order to assess whether the observed fluctuations are indeed due to a measurement uncertainty inherent to the method, the latter quantity can be analytically derived under the hypothesis of Gaussian PSFs $h_{\textrm{in}}$ and $h_{\textrm{out}}$~\cite{bureau_self-portrait_2026}. The result is an intrinsic uncertainty that decreases with the numerical aperture $NA$, the central frequency $f_c$, the echo time $t$, the signal-to-noise ratio $\beta$ and the number of speckle grains considered for local averaging:
\begin{equation}
    \frac{\delta \hat{c}}{c} = \frac{2}{\pi\sqrt{3} } \frac{1}{\beta^{1/2}}\frac{1}{ (N_x N_t)^{1/4}}\frac{1}{ NA^2}\frac{1}{f_c t}
    \label{uncertainty}
\end{equation}
In our experimental conditions (NA=0.5, $\beta$=5, $N_x N_t=16$), the uncertainty calculated with Eq.\eqref{uncertainty} varies between $160$~m.s$^{-1}$ at $z=10$~mm and $46$~m.s$^{-1}$ at $z=35$~mm. This prediction is slightly larger than the variations observed in Fig.~\ref{fig:velocity_map} but we must keep in mind that Eq.~\eqref{uncertainty} has been derived under a Gaussian beam approximation. This expression is therefore only qualitative but it rightly predicts the scaling of uncertainty with the different experimental parameters involved, in particular its decrease with echo time $t$ and effective depth $z$ [Fig.~\ref{fig:velocity_map}(b)].  
\section{Measurement of the shear wave velocity}
In the previous section, the focused reflection matrix was computed for longitudinal waves. In order to determine the mechanical properties of the material, it is necessary to estimate the shear wave velocity.

\begin{figure}[ht]
    \centering
    \includegraphics[width=0.6\linewidth]{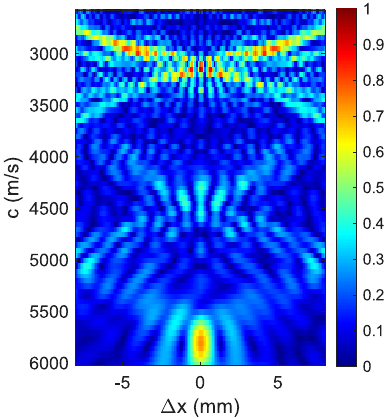}
    \caption{RPSF locally averaged for the central echo time $t=13.8~\mu s$ (corresponding to $z\approx
    40$ mm for longitudinal waves and $z\approx
    22$ mm for shear waves) and in $x=0$ mm on a large range of speed hypotheses.}
    \label{fig:RPSF_wide_range}
\end{figure}

To highlight the contribution of shear waves to the backscattered signals, the RPSF can be evaluated over a wide range of velocity hypotheses. Figure~\ref{fig:RPSF_wide_range} displays the RPSF calculated for an echo time $t=13.8~\mu s$, and for velocities between 2500 m/s and 6000 m/s, from the reflection matrix acquired on Inconel 600. Two dominant maxima can be observed, corresponding, respectively, to longitudinal wave velocity $c_L \approx 5800 m/s$ and shear wave velocity $c_S \approx 3200 m/s$. This result confirms that both wave polarization contribute to the measured backscattered field and suggests that their speeds can be estimated by focusing optimization. Note that  a secondary maximum is observed at an intermediate wave velocity ($c \approx 4450 m/s$) which we attribute to conversion between longitudinal and transverse modes upon reflection. This aspect is beyond the scope of this paper and will require further investigations.

\begin{figure*}[ht]
    \centering
    \includegraphics[width=0.8\linewidth]{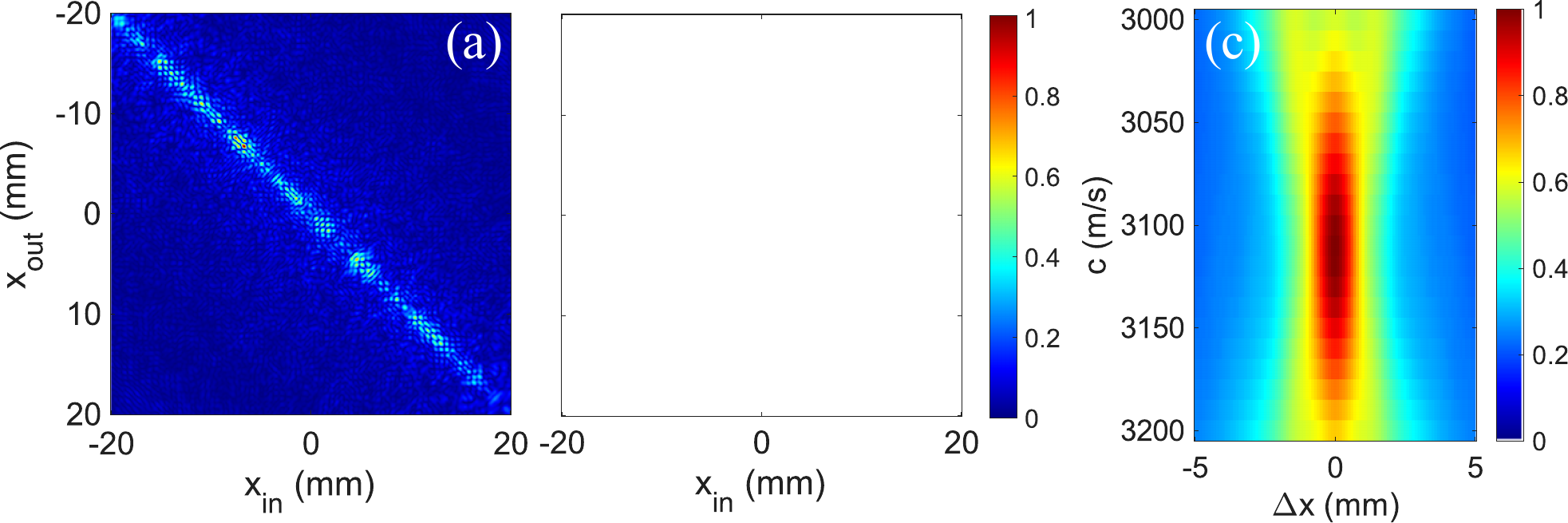}
    \caption{ Focusing of the reflection matrix measured in the Inconel 600 sample: (a) Amplitude of the focused reflection matrix  $ \mathbf{R}_{xx}$  for $c = 3100$~m/s$^{-1}$; (b) amplitude of the focused reflection matrix  adapted for shear waves $\mathbf{R}'_{xx}$ calculated for $c = 3100$~m.s$^{-1}$; (c) locally averaged RPSF calculated from the adapted focused reflection matrix at $x=0$~mm.  The focused reflection matrices and the associated RPSF are obtained at time $t=7.74~\mu s$ and normalized.} 
    \label{fig:mat_shear_nofilt}
\end{figure*}

\begin{figure}[ht]
   \centering
    \includegraphics[width=0.9\linewidth]{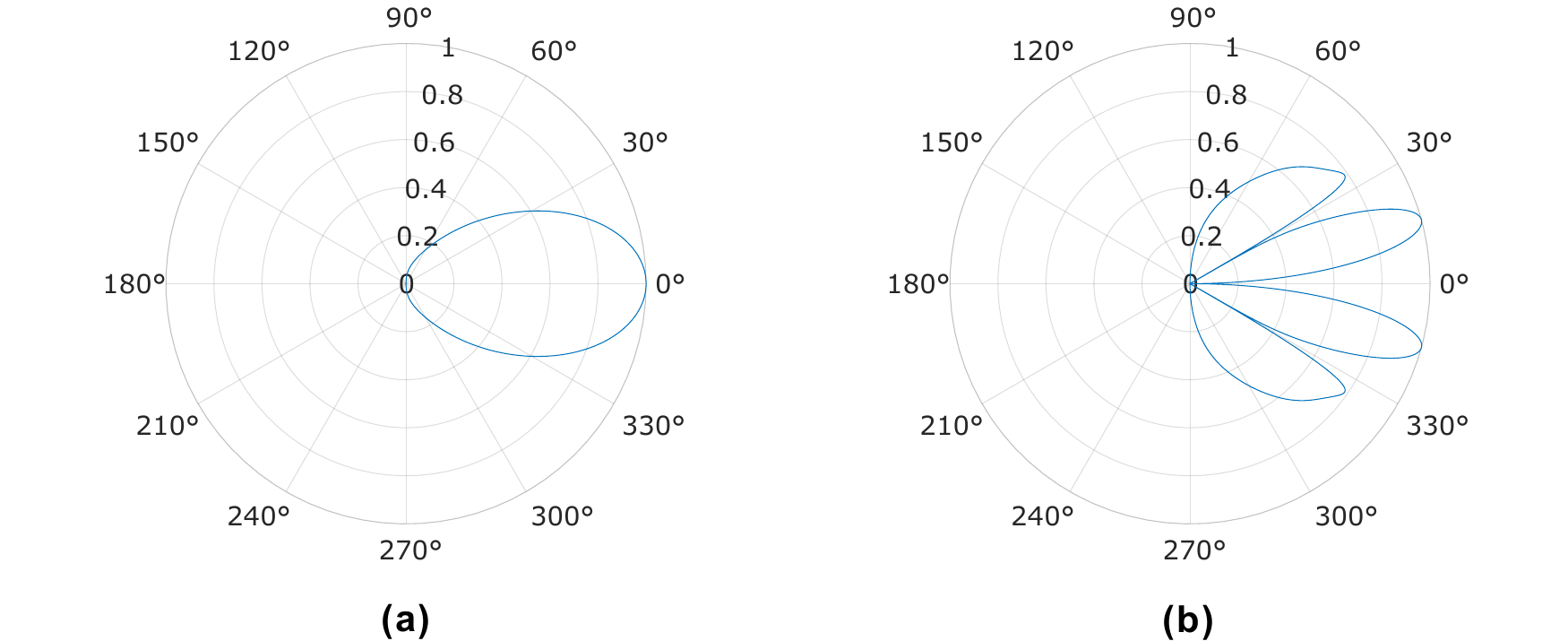}  
\caption{Theoretical (a) longitudinal and (b) shear wave directivities for an element of width $0.4$~mm, a longitudinal speed in the medium of $5848$~m/s, a shear speed of $3121$~m/s, and an emitted  signal composed of a 2-period sinusoid centered at $3.5$~MHz.}
\label{fig:directivity}
\end{figure}

A straightforward approach to evaluate shear wave speeds consists in computing the focused reflection matrix for velocity hypotheses close to the expected shear wave velocity. In this case, the matrix $\mathbf{R}_{xx}(z; c)$ exhibits an enhanced diagonal when the wave velocity model $c$ is close to the shear wave velocity $c_S$, as shown in Fig.~\ref{fig:mat_shear_nofilt}(a).
However, strong interferences appear around the diagonal of the matrix $\mathbf{R}_{xx}(z; c_S)$.  These interference modulate the shear wave RPSF  (Fig.~\ref{fig:RPSF_wide_range}),  preventing a reliable estimation of $c_S$. This effect is related to the directivity of shear wave emission by the transducers. Theoretical directivity patterns, computed from the expressions given in \cite{langenberg_ultrasonic_2017}, are presented in Fig.~\ref{fig:directivity}. 
In contrast to longitudinal waves, shear waves are not efficiently emitted in the normal direction of each element. This results in a lack of low spatial frequencies in the transmitted field. Consequently, the focusing spot is dominated by the interference of contributions originating from both sides of the array, which degrades the RPSF and prevents its proper maximization.

In the following, the focused reflection matrix is adapted to overcome this limitation and enable a robust estimation of the shear wave velocity.
\subsection{Adaptation of the focused reflection matrix to shear waves}
The idea consists in extracting the envelope of the focal spot by separating the contributions from the left and right parts of the array and combining them incoherently. In practice, this is achieved by splitting the Green's function into two contributions corresponding to the left and right parts of the array relative to the lateral position of the focusing point. The beamforming operation of Eq.~\eqref{eq:beamforming} is then performed using two restricted Green’s matrices $\mathbf{G}^l_0$ and $\mathbf{G}^r_0$, whose coefficients are defined as:
\begin{equation}
\begin{aligned}
    G_0^l(u,x,z,f;c) &= G_0(u,x,z,f;c)\,H(x-u), \\
    G_0^r(u,x,z,f;c) &= G_0(u,x,z,f;c)\,H(u-x),
\end{aligned}
\end{equation}
where $H$ denotes the Heaviside function. This definition ensures that only elements located on one side of the focusing point contribute to each Green’s matrix. Using these two Green’s matrices, four focused reflection matrices can be obtained:
\begin{equation}
\label{eq0}
\begin{matrix}
    \mathbf{R}_{xx}^{lr}(z;c) = \sum_f \mathbf{G}_0^{l*}(z,f;c)\times \mathbf{R}_{uu}(f) \times \mathbf{G}_0^{r\dag}(z,f;c) \\
    \mathbf{R}_{xx}^{rl}(z;c) = \sum_f \mathbf{G}_0^{r*}(z,f;c)\times \mathbf{R}_{uu}(f)  \times \mathbf{G}_0^{l\dag}(z,f;c)  \\
     \mathbf{R}_{xx}^{ll}(z;c)= \sum_f \mathbf{G}_0^{l*}(z,f;c)\times \mathbf{R}_{uu}(f)  \times \mathbf{G}_0^{l\dag}(z,f;c) \\
  \mathbf{R}_{xx}^{rr}(z;c) = \sum_f\mathbf{G}_0^{r*}(z,f;c) \times \mathbf{R}_{uu}(f)  \times \mathbf{G}_0^{r\dag}(z,f;c)  \\    
\end{matrix}      
\end{equation}
Note that, in virtue of the reciprocity principle, the first two matrices are transposes of each other: $ \mathbf{R}_{xx}^{lr} \equiv  \mathbf{R}_{xx}^{lr\top}$, where the symbol $\top$ stands for matrix transposition.

\begin{figure}[ht]
    \centering
    \includegraphics[width=0.99\linewidth]{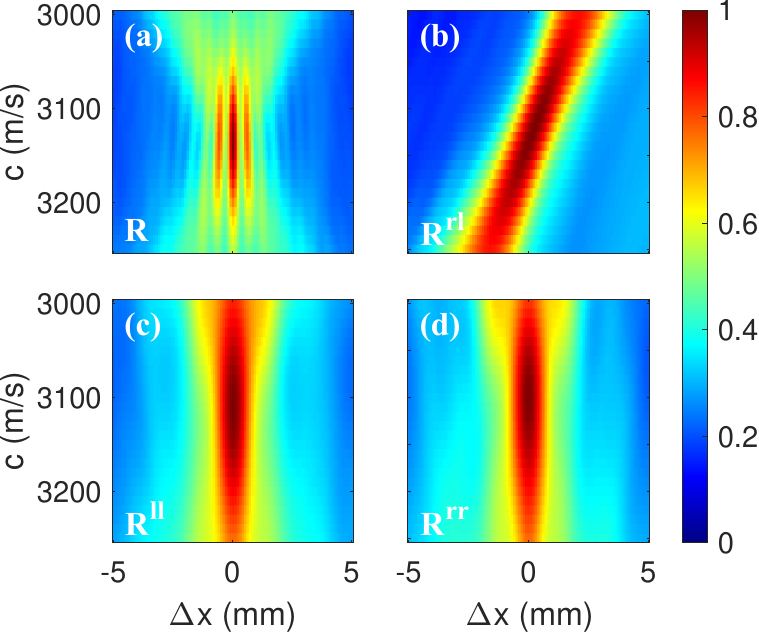}
    \caption{Comparison of the locally averaged RPSFs at time $t=7.74$~$\mu$s and $x=0$~mm. (a) Initial RPSF before spatial filtering, displaying oscillations. (b) RPSF extracted from $\mathbf{R}_{xx}^{rl}$ (which is symmetric to $\mathbf{R}_{xx}^{lr}$). (c) RPSF extracted from $\mathbf{R}_{xx}^{rr}$. (d) RPSF extracted from $\mathbf{R}_{xx}^{ll}$. }
    \label{fig:rpsf_intermediaires}
\end{figure}

To illustrate the benefits of this spatial decomposition, Fig.~\ref{fig:rpsf_intermediaires} displays the RPSF extracted at time $t=7.74$~$\mu$s and $x=0$~mm for the different configurations. Figure~\ref{fig:rpsf_intermediaires}(a) shows the initial RPSF obtained before applying the spatial filter, which exhibits strong oscillations due to phase interferences. In contrast, the RPSFs derived from the restricted matrices $\mathbf{R}_{xx}^{rl}$ [Fig.~\ref{fig:rpsf_intermediaires}(b)], $\mathbf{R}_{xx}^{rr}$ [Fig.~\ref{fig:rpsf_intermediaires}(c)], and $\mathbf{R}_{xx}^{ll}$ [Fig.~\ref{fig:rpsf_intermediaires}(d)] are free of these parasitic oscillations. Moreover, these matrices carry complementary spatial information about the focusing process. 

To reconstruct an optimal focal spot while suppressing phase-related interference, the four matrices defined in Eq.~\eqref{eq0} are incoherently compounded, yielding an interference-free focused reflection matrix :
%To reconstruct an optimal focal spot while avoiding the return of phase-related interferences, these matrices can be incoherently compounded. An interference-free focused reflection matrix is thus computed by incoherent summation of the four matrices 
\begin{equation}
    \mathbf{R}'_{xx} = |\mathbf{R}_{xx}^{lr}| + |\mathbf{R}_{xx}^{rl}| + |\mathbf{R}_{xx}^{ll}| + |\mathbf{R}_{xx}^{rr}|
    \label{eq:Rrr_inc}
\end{equation}
As can be observed by comparing the original matrix $ \mathbf{R}_{xx}$ [Fig.~\ref{fig:mat_shear_nofilt}(a)] and its incoherently-compounded counterpart $ \mathbf{R}'_{xx}$ [Fig.~\ref{fig:mat_shear_nofilt}(b)], this operation removes the interference fringe patterns and gives access to a smoothed RPSF [see comparison between Figs.~\ref{fig:rpsf_intermediaires}(a) and \ref{fig:mat_shear_nofilt}(c)] 

\begin{figure}[h]
    \centering
    \includegraphics[width=0.8\linewidth]{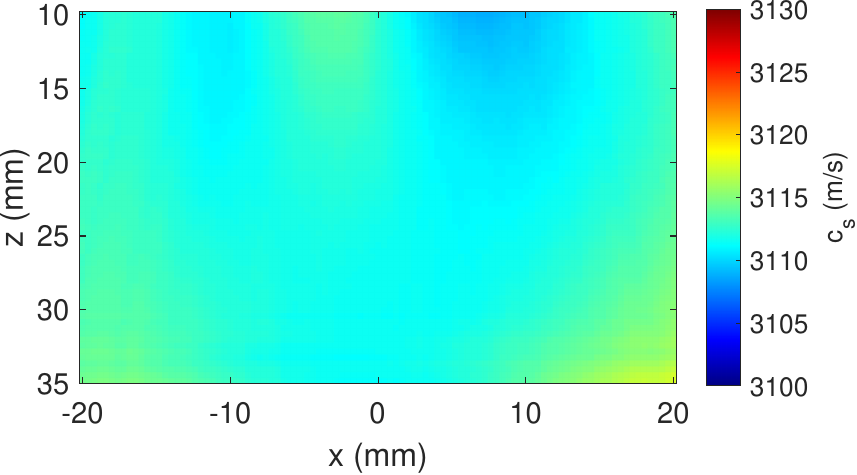} 
    \caption{Shear velocity map of the Inconel 600 sample constructed from the incoherently-compounded reflection matrix {(Eq.~\eqref{eq:Rrr_inc})}. }
    \label{fig:shear_velocity_map}
\end{figure}

In the following, we construct a shear velocity map from $ \mathbf{R}'_{xx}$ using the same approach as for longitudinal waves. The resulting shear velocity distribution is displayed in Fig.~\ref{fig:shear_velocity_map}. 
As before, the sample is expected to be statistically homogeneous, therefore variations observed on this velocity map are attributed to estimation errors rather than material inhomogeneities. Indeed, Eq.~\eqref{uncertainty} still applies and the relative error is in the same order of magnitude as for the longitudinal component. This last statement is confirmed by the measured values of $c_L$ and $c_S$ and their uncertainty reported in Tab.~\ref{tab:elastic_constants} for the steel, Inconel and Ti64 samples. 

This independent mapping of $c_L$ and $c_S$ paves the way towards a non-invasive and in-depth measurement of the mechanical properties of polycrystalline materials, as shown in the following.

\subsection{Computation of the elastic constants}

 %we can deduce the elastic constants of the material from the longitudinal and shear velocity maps,% namely the Young’s modulus ($E$), the shear modulus ($G$), and the Poisson’s ratio ($\nu$).
On the one hand, the Poisson’s ratio $\nu$ can actually be directly computed from the velocity maps using the following equation 
\begin{equation}
    \nu = \frac{1}{2} \left (1-\frac{1}{\kappa^2-1} \right ),
\end{equation}
with $\kappa = c_L/c_S$. On the other hand, the shear modulus $G$ and Young's modulus $E$ are given by:
\begin{equation}
    G =  \rho c_S^2 
\end{equation}
and
\begin{equation}
E = 2G(1+\nu).
\end{equation}
The estimations of Young's modulus $E$ and the shear modulus $G$ 
thus require a prior knowledge of the density $\rho$, which we will assume here to be known and constant. $ \nu $, $G$ and $E$ can therefore be deduced from the prior measurements of $c_L$ and $c_S$. The corresponding values are gathered for the three samples in Table~\ref{tab:elastic_constants}. The low error bars indicate the stability of the measurements and the robustness of the proposed method.

\begin{table}[h]
    \centering
%   \begin{tabular}{|c|c|c|c|c|c|c|}
%        \hline
%        Material &$\rho$ (kg.m$^{-3}$)& $c_l$ (m/s) & $c_s$ (m/s) & $E$ (GPa) & $G$ (GPa) & $\nu$ \\
%        \hline
%        Steel & 7850& $5669\pm9$ & $3100 \pm 9$ & $194 \pm 1$ & $75 \pm 1$ & $0.287 \pm 0.002$\\
%        Inconel &8420 &$5820\pm 7$ & $3112 \pm 1$ & $213 \pm 1$ & $82 \pm 1$ & $0.3 \pm 0.001$ \\
%        Ti64 &4430 &$6161 \pm 12$ & $3176 \pm 7$ & $118 \pm 1$ & $45 \pm 1$ & $0.319 \pm 0.002$ \\
%        \hline
%    \end{tabular}
    \begin{tabular}{|c|c|c|c|}
\hline
Material & Steel & Inconel & Ti64 \\
\hline
$\rho$ (kg.m$^{-3}$) & 7850 & 8420 & 4430 \\
\hline
$c_L$ (m/s) & $5669\pm9$ & $5820\pm7$ & $6161\pm12$ \\
$c_S$ (m/s) & $3100\pm9$ & $3112\pm1$ & $3176\pm7$ \\
$\nu$ & $0.287\pm0.002$ & $0.3\pm0.001$ & $0.319\pm0.002$ \\
$E$ (GPa) & $194\pm1$ & $213\pm1$ & $118\pm1$ \\
$G$ (GPa) & $75\pm1$ & $82\pm1$ & $45\pm1$ \\
\hline
\end{tabular}
    \caption{Measured wave velocities and estimated elastic constants of the three polycrystalline materials.} 
    \label{tab:elastic_constants}
\end{table}

However, a key limitation remains: without a ground truth reference, it is challenging to ensure that the estimated values are indeed the true elastic constants of the materials. In particular, the assumption of a constant density $\rho$ may introduce biases in $E$ and $G$. Further validation using independent reference measurements would be necessary to confirm the accuracy of the method.

\section{Validation Through Numerical Simulations}

%To address the limitations of the method,
To assess the validity of our approach, we perform simulations of samples with controlled and homogeneous effective velocities and densities. Since the method relies on echoes backscattered by the microstructure, an appropriate strategy is needed for their simulation. To this end, we used the 2D finite-difference time-domain simulation software Simsonic2D \cite{laugier_numerical_2011}.  

A fully realistic microstructure simulation would be computationally expensive, as computing the full reflection matrix requires as many simulations as there are array elements. Instead, we adopt a simplified approach in which the medium is modeled as a homogeneous elastic solid containing randomly distributed point-like scatterers that mimic wave interactions with the microstructure. These scatterers share the same elastic constants as the host material but differ in density, thereby introducing local impedance fluctuations.

Two types of scatterers are considered: one with a higher density and one with a lower density. Their densities are chosen such that the associated wave velocities are symmetrically distributed around the reference velocity of the bulk material. This ensures that the mean velocity remains unchanged while introducing controlled heterogeneity. The simulations are performed using material properties representative of Inconel 600. The main simulation parameters, including spatial and temporal sampling, are summarized in Table~\ref{tab:simu_param}.

\begin{table}[htbp]
  \centering
  \begin{tabular}{|c|c|}
    \hline
    Parameter                          & Value \\
    \hline
    Domain size (x)                    & 70 mm \\
    Domain size (z)                    & 110 mm \\
    Spatial step $\Delta x = \Delta z$ & 50 $\mu$m \\
    Simulation duration $T$            & 65 $\mu$s \\
    Probe                             & 128-element linear array \\
    Pitch                             & 0.4 mm \\
    Element width                     & 0.4 mm \\
    Central frequency $f_0$            & 3.5 MHz \\
    Scatterer fraction                & 8 \% \\
    Impedance contrasts               & $\pm 2.5\%$ \\
    \hline
  \end{tabular}
  \caption{Summary of the numerical simulation parameters.}
  \label{tab:simu_param}
\end{table}

The velocities and elastics constants estimated for this simulation are compared in Table~\ref{tab:simulation_comparison}, where we report the measured Poisson's ratio, Young's and shear moduli, as well as the relative differences with the true simulated value.  

\begin{table}[h]
    \centering
    \begin{tabular}{|c|c|c|c|}
        \hline
        Property & Estimated & Ground-truth & Error (\%) \\
        \hline
        $c_L$ (m.s$^{-1}$) & $5846$ & $5842$ & $0.1$ \\
        $c_S$ (m.s$^{-1}$) & $3102$ & $3117$ & $0.5$ \\
        $\nu$ & $0.304$ & $0.301$ & $1$ \\
        $E$ (GPa) & $212.4$ & $214.2$ & $0.8$ \\
        $G$ (GPa) & $81.5$ & $82.3$ & $1$ \\

        \hline
    \end{tabular}
    \caption{Comparison between estimated and ground-truth values of elastic constants for the simulation of Inconel 600. The error represents the deviation between the estimated and the ground truth values}.
    \label{tab:simulation_comparison}
\end{table}

These results demonstrate the excellent accuracy of the proposed method in a controlled case where the ground truth is known. They confirm the robustness of the approach applied to media with homogeneous effective velocities.  
Further investigations are required to assess the performance of the method in materials displaying heterogeneous mechanical properties.

\section{Simulation of a Heterogeneous Wave velocity Distribution Medium}

As demonstrated in the previous section, the proposed method yields satisfactory results for homogeneous effective wave velocity distribution. However, due to the lack of available metallic samples exhibiting significant and well-controlled macroscopic inhomogeneities, the heterogeneous case has not yet been investigated experimentally.

\begin{figure}[h]
    \centering
    \includegraphics[width=0.99\linewidth]{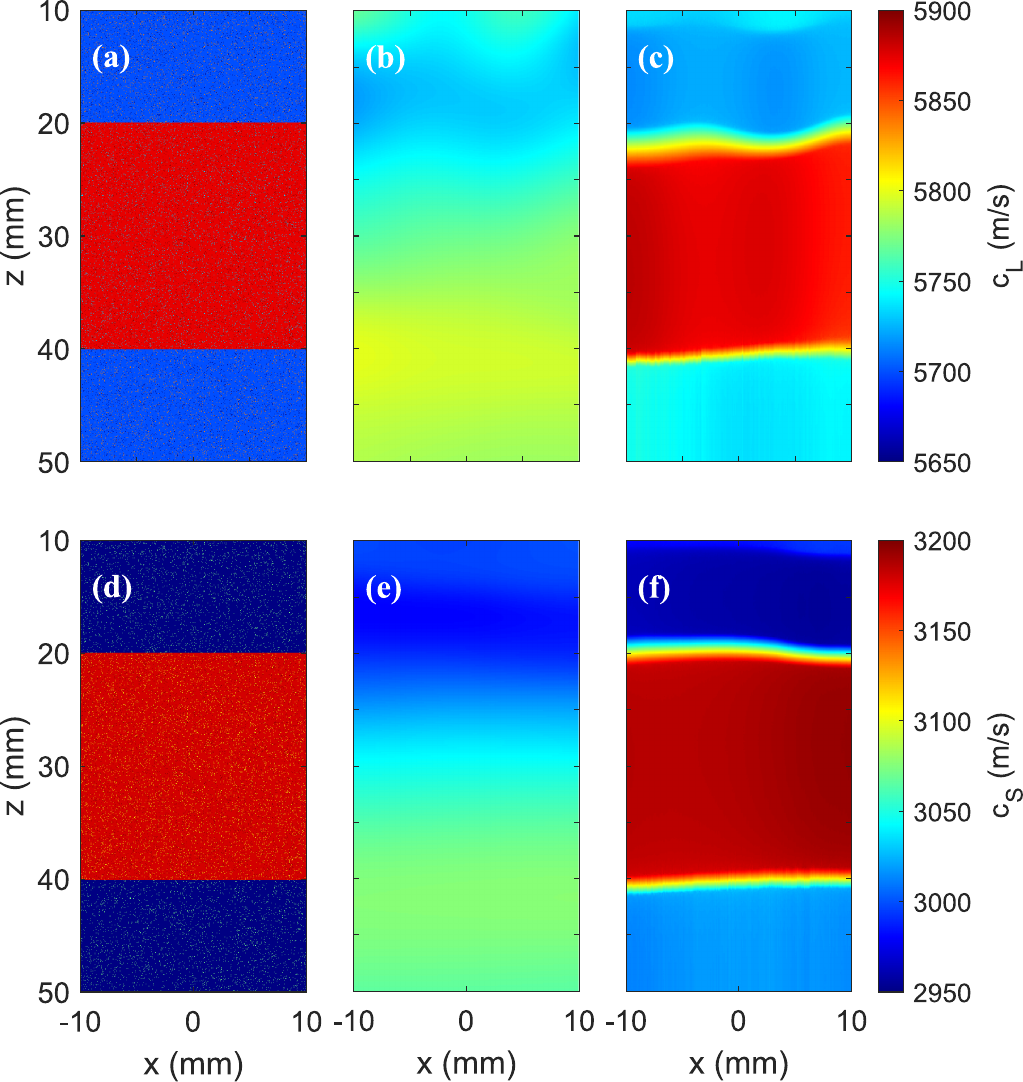} 
    \caption{Longitudinal (a) and shear (d) velocity maps of the simulated medium, optimal longitudinal (b) and shear (e) velocity maps and local longitudinal (c) and shear (f) velocity maps. }
    \label{fig:layered_velocity_maps}
\end{figure}

To assess the potential of the method in this kind of situation, we use the same simulation framework as above, with elastic constants corresponding to those measured in the previous section for homogeneous materials. More precisely, a three layer medium is simulated as depicted in Fig.~\ref{fig:layered_velocity_maps}(a) and (d). The configuration consists of an Inconel–steel–Inconel stack. Steel is selected as the intermediate layer because it is a widely used industrial material whose wave velocities are close to those of Inconel.

It is important to note that, in the simulations,  the interfaces are perfectly flat, leading to strong and coherent specular reflections.  In contrast, the microstructure is modeled using sparse point-like scatterers, which produce weaker distributed backscattering than in real materials. As a result, interface reflections dominate the simulated signals therefore masking the contribution of microstructural backscattering. To mitigate this effect, the  signals obtained from a reference simulation without scatterers are subtracted, thereby removing the specular reflections and isolating the contribution of the scattering by the microsructure.  

The optimal longitudinal and shear velocity maps, $\hat{c}(x,z)$, obtained for this model medium are shown in Fig.~\ref{fig:layered_velocity_maps}(b) and (e). 

%\sout{While the optimal velocity maps capture the overall variations, they do not adequately represent local velocities.  We thus propose to retrieve local velocities from the optimal ones by solving an inverse problem.}
They are far from the ground truth distributions displayed in Fig.~\ref{fig:layered_velocity_maps}(a) and (d).
Indeed, the optimized wave velocity $\hat{c}$ is not an estimator for the local speed-of-sound but for
the inverse of the mean slowness $\bar{s}(x, t)$, averaged between the probe surface and the focusing point $(x, t)$~\cite{bureau_self-portrait_2026}.
Under a geometrical acoustics framework and assuming near-vertical propagation (paraxial approximation), the  mean slowness$\overline{s}(x,t)$ at a given lateral position $x$ can be expressed as the depth-average of the local slowness $1/c(x,z)$~\cite{jakovljevic_local_2018}:
\begin{equation}
\label{eq1}
    \overline{s}(x,t)=\frac{1}{z_t(x)}\int_{0}^{z_t(x)} \frac{dz}{c(x,z)},
\end{equation}
where $z_t(x)= \hat{c}(x,t) t / 2$ is the depth of scatterers contributing to echoes received at time $t$.

However, this formulation does not account for the spatial averaging introduced in the RPSF estimation. As described in the previous section, the RPSF is computed using a Gaussian weighting window of temporal extent $\sigma_{t}$, which defines the size of the averaging region along the propagation direction. This spatial averaging induces an effective smoothing of the mean slowness.
%Expressed in the time domain, this corresponds to a convolution along the echo time $t$ with a Gaussian kernel of standard deviation $\sigma_t = \sigma_z / c_0$. 
As a consequence, the measured quantity corresponds to a filtered version of $\overline{s}$ rather than to $\overline{s}$ itself.

A more realistic forward model should therefore incorporate this effect as:
\begin{equation}
\label{eq2}
    \hat{c}^{-1}(x,t)=\overline{s}(x,t) \stackrel{t}{\circledast} \frac{1}{\sqrt{2\pi}\sigma_t}\exp\left(-\frac{t^2}{2\sigma_t^2}\right),
\end{equation}
The local velocity $c(x,z)$ is then estimated by solving the corresponding inverse problem. The optimization is performed using a total variation regularization with an $\ell_1$-norm \cite{cherkaoui_learning_2020}, which promotes piecewise-constant profiles and is well suited to layered media. 

The resulting local velocity maps are shown in Fig.~\ref{fig:layered_velocity_maps}(c) and (f), for both longitudinal and shear waves.

These maps clearly reveal the three-layer structure of the medium and are in good agreement with the ground truth distributions (Fig.~\ref{fig:layered_velocity_maps}(a) and (d)). The estimated velocities, gathered in Table~\ref{tab:velocity_comparison_layers}, are close to the simulated values. 
\begin{table}[h]
\centering
\begin{tabular}{|c|c|c|c|c|}
\hline
Layer & $c_L$ (est.) & $c_L$ (sim.) & $c_S$ (est.) & $c_S$ (sim.) \\
\hline
1 & 5720 & 5700 & 2960 & 3000 \\
2 & 5876 & 5874 & 3189 & 3179 \\
3 & 5736 & 5700 & 3019 & 3000 \\
\hline
\end{tabular}
\caption{Comparison between estimated and simulated longitudinal and shear velocities (in m/s).}
\label{tab:velocity_comparison_layers}
\end{table}
The interfaces between the layers are also well retrieved, appearing at depths close to the ground truth (20~mm and 40~mm), with deviations smaller than 3~mm. The transitions between layers are sharp, occurring over distances smaller than 3~mm, which is less than two wavelengths. This observation highlights the good axial resolution of the method.

This numerical simulation corresponds to a relatively simple layered configuration, without lateral variations, which is well suited to the proposed approach. In practice, the method relies on a homogeneous propagation model to compute the focused reflection matrix, which may limit its accuracy in the presence of strong lateral velocity variations~\cite{bureau_self-portrait_2026}.

\begin{figure}[h!]
    \centering
    \includegraphics[width=0.99\linewidth]{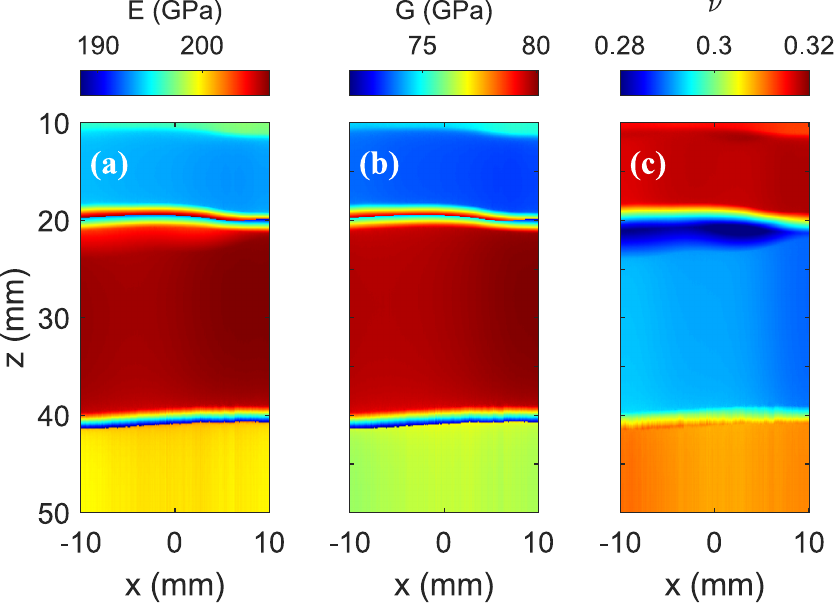} 
    \caption{Local Young's modulus (a), shear modulus (b) and Poisson's ratio (c) maps obtained from the longitudinal and shear velocity maps.}
    \label{fig:cste_elastiques}
\end{figure}

As in Section 3.2, local elastic constants can be derived from the longitudinal and shear velocity maps. The Poisson’s ratio $\nu$ is obtained from the ratio of the two velocities, while the Young’s modulus $E$ and shear modulus $G$ are computed by assuming a known density. In practice, this density must be estimated or measured; in the present simulation, the average density of each layer is used ($\rho=8470$~kg/m$^3$ for Inconel and $\rho=7850$~kg/m$^3$ for steel). The resulting maps are presented in Fig.~\ref{fig:cste_elastiques}.

The three layers are clearly identified in all maps, with values consistent with the expected elastic properties. Slight discontinuities are observed at the interfaces, which are related to small mismatches between their positions in longitudinal and shear velocity maps. Nevertheless, the estimated values remain in very good agreement with the simulated ones, as summarized in Table~\ref{tab:elastic_constants_multilayer}.

\begin{table}[h]
\centering
\begin{tabular}{|c|c|c|c|c|}
\hline
Layer & $E$ (est.) & $E$ (sim.) & $G$ (est.) & $G$ (sim.) \\
\hline
1 & 193.4 & 197.6 & 73.4 & 75.5 \\
2 & 206.0 & 205.1 & 79.8 & 79.3 \\
3 & 200.1 & 197.6 & 76.5 & 75.5 \\
\hline
\end{tabular}
\caption{Comparison between estimated and simulated Young’s and shear moduli (in GPa).}
\label{tab:elastic_constants_multilayer}
\end{table}

\begin{table}[h]
\centering
\begin{tabular}{|c|c|c|}
\hline
Layer & $\nu$ (est.) & $\nu$ (sim.) \\
\hline
1 & 0.317 & 0.308 \\
2 & 0.291 & 0.293 \\
3 & 0.308 & 0.308 \\
\hline
\end{tabular}
\caption{Comparison between estimated and simulated Poisson’s ratios.}
\label{tab:poisson}
\end{table}

These results demonstrate that the proposed approach enables not only the reconstruction of velocity maps, but also the quantitative estimation of local elastic properties within the bulk of the material. This is particularly valuable, as it provides direct access to mechanical parameters that are typically difficult to measure remotely and non-destructively. Such volumetric mapping of elastic constants opens promising perspectives for material characterization and the detection of mechanically critical regions in complex structures.

\section{Conclusion}

In this study, we develop a reflection matrix method for mapping the elastic constants of a material using reflection mapping of the elastic longitudinal and shear velocities. The proposed approach has demonstrated remarkable performance in polycrystalline media, providing reliable and accurate estimates of elastic constants. Precision remains high, with measurement deviations lower than $2~\%$. One  key advantage of this method is that it only requires access to a single surface of the sample. Another advantage is that it can be used with minimal prior knowledge on the material's properties.

However, this approach has certain limitations: lateral velocity variations are not well captured, and the axial resolution remains limited to a few wavelengths.  These limitations could potentially be mitigated by integrating complementary techniques from the medical ultrasound domain, such as CUTE~\cite{jaeger_computed_2015,Staehli2020}, which offers better lateral resolution but strongly depends on the initial velocity assumption.

Nevertheless, deriving a wave velocity map from a mean slowness ~\cite{Ali2022,jakovljevic_local_2018} or an integrated phase shift like in CUTE~\cite{jaeger_computed_2015,Staehli2020} often requires a straight-ray approximation that neglects refraction phenomena. Interestingly, matrix imaging and its intrinsic focusing metrics have been recently exploited for optimizing a forward model accounting for refraction and forward multiple scattering in medical ultrasound~\cite{HeriardDubreuil2026}. The extension of this approach to elastic waves and polycrystalline materials is an evident and promising perspective of this work.

More fundamentally, another important limit of our method and others is their range of validity with respect to the scattering regime induced by the microstructure. If the medium is either too weakly or too strongly scattering, the single back-scattering approximation, on which this work is based, will not hold. The penetration depth limit is of the order of the scattering mean free path, the mean distance between two scattering events, which is also the typical distance beyond which the multiple scattering background would become predominant~\cite{Goicoechea2024}.

Despite its limits, the proposed method opens important perspectives for volumetric characterization of materials and beyond. In particular, it may be used to enhance the quality of reflectivity images by compensating for aberrations induced by velocity variations within the material. Interestingly, the RPSF concept can be used to study mode conversion between the longitudinal and shear elastic components as waves propagate through the sample (Fig.~\ref{fig:RPSF_wide_range}). This feature can be a relevant observable for the detection of micro-textured regions in Titanium alloys~\cite{hashimoto_effects_2023}. The level of mode conversion is also particularly important to monitor the presence of fluids in seismology~\cite{Kawakatsu2009}, a field in which reflection matrix imaging has already shown its strength~\cite{Giraudat2024}. Thus the proposed method, with potential extensions that could further enhance its applicability and accuracy, provides a powerful tool for characterizing textured materials with elastic waves, whether it be for non destructive testing with ultrasound or seismic monitoring at much larger scale. 

\section*{CRediT authorship contribution statement}

\textbf{Gatien Clement:} Writing – original draft, Conceptualization, Formal analysis, Visualization, Investigation, Data curation, Writing – review \& editing. 
\textbf{Alexandre Aubry:} Conceptualization, Supervision, Writing – review \& editing. 
\textbf{C\'ecile Br\"utt:} Funding acquisition, Writing – review \& editing. 
\textbf{Beno\^it G\'erardin:} Funding acquisition, Writing – review \& editing. 
\textbf{Claire Prada:} Conceptualization, Supervision, Writing – review \& editing.

\section*{Declaration of competing interests}

A.A. is an inventor on a patent related to this work (no. EP4731998, filed in June 2024). G.C., A.A., C.B., B.G., and C.P. are inventors on another patent related to this work (no. FR3169218A1, filed in December 2024). C.B. and B.G. are employees of SAFRAN company. All authors declare that they have no other competing interests.

\section*{Acknowledgments} 

This work has received support under the program “Investissements d’Avenir” launched by the French Government, LABEX WIFI (Laboratory of Excellence within the French Program Investments for the Future, ANR-10-LABX24 and ANR-10-IDEX-0001–02 PSL). The authors thank Arnaud Derode for allowing the use of a Verasonic multi-channel electronic device as well as the Inconel 600 sample. They also thank Baptiste Hériard-Dubreuil for his help with the implementation of the optimization method and for his valuable advice on regularization.

\section*{Data availability}

The data that support the findings of this study are available from the corresponding author upon reasonable request.

\bibliographystyle{elsarticle-num-names} 
\bibliography{Article}

@article{jakovljevic_local_2018,
	title = {Local {Speed} of {Sound} {Estimation} in {Tissue} {Using} {Pulse}-{Echo} {Ultrasound}: {Model}-{Based} {Approach}},
	volume = {144},
	issn = {0001-4966},
	shorttitle = {Local {Speed} of {Sound} {Estimation} in {Tissue} {Using} {Pulse}-{Echo} {Ultrasound}},
	doi = {10.1121/1.5043402},
	number = {1},
	journal = {The Journal of the Acoustical Society of America},
	author = {Jakovljevic, Marko and Hsieh, Scott and Ali, Rehman and Chau Loo Kung, Gustavo and Hyun, Dongwoon and Dahl, Jeremy J.},
	month = jul,
	year = {2018},
	pages = {254--266},
}

@article{lambert_reflection_2020,
	title = {Reflection {Matrix} {Approach} for {Quantitative} {Imaging} of {Scattering} {Media}},
	volume = {10},
	doi = {10.1103/PhysRevX.10.021048},
	number = {2},
	journal = {Physical Review X},
	author = {Lambert, William and Cobus, Laura A. and Couade, Mathieu and Fink, Mathias and Aubry, Alexandre},
	month = jun,
	year = {2020},
	pages = {021048},
}

@article{baelde_effect_2018,
	title = {Effect of {Microstructural} {Elongation} on {Backscattered} {Field}: {Intensity} {Measurement} and {Multiple} {Scattering} {Estimation} with a {Linear} {Transducer} {Array}},
	volume = {82},
	issn = {0041-624X},
	shorttitle = {Effect of {Microstructural} {Elongation} on {Backscattered} {Field}},
	doi = {10.1016/j.ultras.2017.09.006},
	journal = {Ultrasonics},
	author = {Baelde, Aurélien and Laurent, Jérôme and Millien, Pierre and Coulette, Richard and Khalifa, Warida Ben and Jenson, Frédéric and Sun, Fan and Fink, Mathias and Prada, Claire},
	month = jan,
	year = {2018},
	pages = {379--389},
}

@article{lan_experimental_2014,
	title = {Experimental and computational studies of ultrasound wave propagation in hexagonal close-packed polycrystals for texture detection},
	volume = {63},
	issn = {13596454},
	url = {https://linkinghub.elsevier.com/retrieve/pii/S1359645413007635},
	doi = {10.1016/j.actamat.2013.10.012},
	language = {en},
	urldate = {2022-11-16},
	journal = {Acta Materialia},
	author = {Lan, B. and Lowe, M. and Dunne, F.P.E.},
	month = jan,
	year = {2014},
	pages = {107--122},
}

@article{holmes_post-processing_2005,
	title = {Post-processing of the full matrix of ultrasonic transmit–receive array data for non-destructive evaluation},
	volume = {38},
	copyright = {https://www.elsevier.com/tdm/userlicense/1.0/},
	issn = {09638695},
	url = {https://linkinghub.elsevier.com/retrieve/pii/S0963869505000721},
	doi = {10.1016/j.ndteint.2005.04.002},
	language = {en},
	number = {8},
	urldate = {2025-01-27},
	journal = {NDT \& E International},
	author = {Holmes, Caroline and Drinkwater, Bruce W. and Wilcox, Paul D.},
	month = dec,
	year = {2005},
	pages = {701--711},
}

@incollection{laugier_numerical_2011,
	address = {Dordrecht},
	title = {Numerical {Methods} for {Ultrasonic} {Bone} {Characterization}},
	isbn = {978-94-007-0016-1 978-94-007-0017-8},
	url = {https://link.springer.com/10.1007/978-94-007-0017-8_8},
	doi = {10.1007/978-94-007-0017-8_8},
	language = {en},
	urldate = {2025-01-27},
	booktitle = {Bone {Quantitative} {Ultrasound}},
	publisher = {Springer Netherlands},
	author = {Bossy, Emmanuel and Grimal, Quentin},
	editor = {Laugier, Pascal and Haïat, Guillaume},
	year = {2011},
	pages = {181--228},
}

@article{lambert_ultrasound_2022,
	title = {Ultrasound {Matrix} {Imaging}—{Part} {II}: {The} {Distortion} {Matrix} for {Aberration} {Correction} {Over} {Multiple} {Isoplanatic} {Patches}},
	volume = {41},
	copyright = {https://ieeexplore.ieee.org/Xplorehelp/downloads/license-information/IEEE.html},
	issn = {0278-0062, 1558-254X},
	shorttitle = {Ultrasound {Matrix} {Imaging}—{Part} {II}},
	url = {https://ieeexplore.ieee.org/document/9858891/},
	doi = {10.1109/TMI.2022.3199483},
	number = {12},
	urldate = {2025-01-27},
	journal = {IEEE Transactions on Medical Imaging},
	author = {Lambert, William and Cobus, Laura A. and Robin, Justine and Fink, Mathias and Aubry, Alexandre},
	month = dec,
	year = {2022},
	pages = {3921--3938},
}

@book{langenberg_ultrasonic_2017,
	address = {Boca Raton},
	title = {Ultrasonic nondestructive testing of materials: theoretical foundations},
	isbn = {978-1-138-07596-2},
	shorttitle = {Ultrasonic nondestructive testing of materials},
	language = {eng},
	publisher = {CRC Press, Taylor et Francis},
	author = {Langenberg, Karl-Jörg and Marklein, René and Meyer, Klaus},
	year = {2017},
	note = {OCLC: 1019892978},
}

@article{leisure_resonant_1997,
	title = {Resonant ultrasound spectroscopy},
	volume = {9},
	issn = {0953-8984, 1361-648X},
	url = {https://iopscience.iop.org/article/10.1088/0953-8984/9/28/002},
	doi = {10.1088/0953-8984/9/28/002},
	number = {28},
	urldate = {2025-02-12},
	journal = {Journal of Physics: Condensed Matter},
	author = {Leisure, R G and Willis, F A},
	month = jul,
	year = {1997},
	pages = {6001--6029},
}

@book{aki_quantitative_2009,
	address = {Mill Valley, California New York},
	edition = {2. edition, corrected printing},
	title = {Quantitative seismology},
	isbn = {978-1-891389-63-4},
	language = {eng},
	publisher = {University Science Books},
	author = {Aki, Keiiti and Richards, Paul G.},
	year = {2009},
}

@book{grechka_encyclopedia_2014,
	title = {Encyclopedia of {Exploration} {Geophysics}},
	isbn = {978-1-56080-301-0 978-1-56080-302-7},
	url = {https://library.seg.org/doi/book/10.1190/1.9781560803027},
	doi = {10.1190/1.9781560803027},
	language = {en},
	urldate = {2025-02-12},
	publisher = {Society of Exploration Geophysicists},
	editor = {Grechka, Vladimir and Wapenaar, Kees},
	month = jan,
	year = {2014},
}

@article{anderson_direct_1998,
	title = {The direct estimation of sound speed using pulse–echo ultrasound},
	volume = {104},
	issn = {0001-4966, 1520-8524},
	url = {https://pubs.aip.org/jasa/article/104/5/3099/560645/The-direct-estimation-of-sound-speed-using-pulse},
	doi = {10.1121/1.423889},
	language = {en},
	number = {5},
	urldate = {2025-02-12},
	journal = {The Journal of the Acoustical Society of America},
	author = {Anderson, Martin E. and Trahey, Gregg E.},
	month = nov,
	year = {1998},
	pages = {3099--3106},
}

@article{feigin_deep_2020,
	title = {A {Deep} {Learning} {Framework} for {Single}-{Sided} {Sound} {Speed} {Inversion} in {Medical} {Ultrasound}},
	volume = {67},
	copyright = {https://ieeexplore.ieee.org/Xplorehelp/downloads/license-information/IEEE.html},
	issn = {0018-9294, 1558-2531},
	url = {https://ieeexplore.ieee.org/document/8772124/},
	doi = {10.1109/TBME.2019.2931195},
	number = {4},
	urldate = {2025-03-03},
	journal = {IEEE Transactions on Biomedical Engineering},
	author = {Feigin, Micha and Freedman, Daniel and Anthony, Brian W.},
	month = apr,
	year = {2020},
	pages = {1142--1151},
}

@inproceedings{heller_speed--sound_2023,
	address = {Montreal, QC, Canada},
	title = {Speed-of-{Sound} {Reconstruction} with {Deep} {Neural} {Networks} in {Pulse}-{Echo} {Mode}: {Coherency}- {Vs} {RF}-{Data}-{Based} {Approach}},
	copyright = {https://doi.org/10.15223/policy-029},
	isbn = {979-8-3503-4645-9},
	shorttitle = {Speed-of-{Sound} {Reconstruction} with {Deep} {Neural} {Networks} in {Pulse}-{Echo} {Mode}},
	url = {https://ieeexplore.ieee.org/document/10307895/},
	doi = {10.1109/IUS51837.2023.10307895},
	urldate = {2025-03-03},
	booktitle = {2023 {IEEE} {International} {Ultrasonics} {Symposium} ({IUS})},
	publisher = {IEEE},
	author = {Heller, Marvin and Schmitz, Georg},
	month = sep,
	year = {2023},
	pages = {1--4},
}

@article{jaeger_computed_2015,
	title = {Computed {Ultrasound} {Tomography} in {Echo} {Mode} for {Imaging} {Speed} of {Sound} {Using} {Pulse}-{Echo} {Sonography}: {Proof} of {Principle}},
	volume = {41},
	issn = {03015629},
	shorttitle = {Computed {Ultrasound} {Tomography} in {Echo} {Mode} for {Imaging} {Speed} of {Sound} {Using} {Pulse}-{Echo} {Sonography}},
	url = {https://linkinghub.elsevier.com/retrieve/pii/S0301562914003500},
	doi = {10.1016/j.ultrasmedbio.2014.05.019},
	language = {en},
	number = {1},
	urldate = {2025-03-03},
	journal = {Ultrasound in Medicine \& Biology},
	author = {Jaeger, Michael and Held, Gerrit and Peeters, Sara and Preisser, Stefan and Grünig, Michael and Frenz, Martin},
	month = jan,
	year = {2015},
	pages = {235--250},
}

@article{cen_inclusions_2019,
	title = {Inclusions in melting process of titanium and titanium alloys},
	volume = {16},
	issn = {1672-6421, 2365-9459},
	url = {http://link.springer.com/10.1007/s41230-019-9046-1},
	doi = {10.1007/s41230-019-9046-1},
	language = {en},
	number = {4},
	urldate = {2025-03-25},
	journal = {China Foundry},
	author = {Cen, Meng-jiang and Liu, Yuan and Chen, Xiang and Zhang, Hua-wei and Li, Yan-xiang},
	month = jul,
	year = {2019},
	pages = {223--231},
}

@misc{bureau_self-portrait_2026,
	title = {Self-{Portrait} of the {Focusing} {Process} in {Speckle}: {II}. {Gouy} {Phase} {Shift} for {Defocus} {Correction} and {Pixel} {Depth} {Reassignment}},
	shorttitle = {Self-{Portrait} of the {Focusing} {Process} in {Speckle}},
	url = {http://arxiv.org/abs/2409.13901},
	doi = {10.48550/arXiv.2409.13901},
	language = {en},
	urldate = {2026-04-13},
	publisher = {arXiv},
	author = {Bureau, Flavien and Brenner, Emma and Martiartu, Naiara Korta and Giraudat, Elsa and Ber, Arthur Le and Lambert, William and Carmier, Louis and Guibal, Aymeric and Fink, Mathias and Aubry, Alexandre},
	month = feb,
	year = {2026},
	note = {arXiv:2409.13901 [physics]},
}

@article{du_burck_attenuation_2026,
	title = {Attenuation and dispersion of coherent {Rayleigh} and head waves on the surface of a polycrystal},
	volume = {159},
	issn = {1520-8524},
	url = {https://pubs.aip.org/jasa/article/159/1/955/3378305/Attenuation-and-dispersion-of-coherent-Rayleigh},
	doi = {10.1121/10.0042276},
	language = {en},
	number = {1},
	urldate = {2026-04-13},
	journal = {The Journal of the Acoustical Society of America},
	author = {Du Burck, Clément and Derode, Arnaud},
	month = jan,
	year = {2026},
	pages = {955--973},
}

@article{shahjahan_comparison_2014,
	title = {Comparison between experimental and 2-{D} numerical studies of multiple scattering in {Inconel600}® by means of array probes},
	volume = {54},
	issn = {0041624X},
	url = {https://linkinghub.elsevier.com/retrieve/pii/S0041624X13001844},
	doi = {10.1016/j.ultras.2013.06.012},
	language = {en},
	number = {1},
	urldate = {2026-04-13},
	journal = {Ultrasonics},
	author = {Shahjahan, S. and Rupin, F. and Aubry, A. and Chassignole, B. and Fouquet, T. and Derode, A.},
	month = jan,
	year = {2014},
	pages = {358--367},
}

@Misc{HeriardDubreuil2026,
  author    = {Hériard-Dubreuil, Baptiste and Brenner, Emma and Rio, Benjamin and Lambert, William and Chamming's, Foucauld and Fink, Mathias and Aubry, Alexandre},
  title     = {Physics-Based Learning of the Wave Speed Landscape in Complex Media},
  year      = {2026},
  copyright = {Creative Commons Attribution Non Commercial No Derivatives 4.0 International},
  doi       = {10.48550/ARXIV.2602.03281},
  publisher = {arXiv},
}

@Article{Simson2026,
  author    = {Simson, Walter and Zhuang, Louise and Frey, Benjamin N. and Sanabria, Sergio J. and Dahl, Jeremy J. and Hyun, Dongwoon},
  journal   = {IEEE Transactions on Medical Imaging},
  title     = {Ultrasound Autofocusing: Common Midpoint Phase Error Optimization via Differentiable Beamforming},
  year      = {2026},
  issn      = {1558-254X},
  month     = Feb,
  number    = {2},
  pages     = {681--692},
  volume    = {45},
  doi       = {10.1109/tmi.2025.3607875},
  publisher = {Institute of Electrical and Electronics Engineers (IEEE)},
}

@Article{Staehli2020,
  author    = {Stähli, Patrick and Kuriakose, Maju and Frenz, Martin and Jaeger, Michael},
  journal   = {Ultrasonics},
  title     = {Improved forward model for quantitative pulse-echo speed-of-sound imaging},
  year      = {2020},
  issn      = {0041-624X},
  month     = Dec,
  pages     = {106168},
  volume    = {108},
  doi       = {10.1016/j.ultras.2020.106168},
  publisher = {Elsevier BV},
}

@Article{Ali2022,
  author    = {Ali, Rehman and Telichko, Arsenii V. and Wang, Huaijun and Sukumar, Uday K. and Vilches-Moure, Jose G. and Paulmurugan, Ramasamy and Dahl, Jeremy J.},
  journal   = {IEEE Transactions on Ultrasonics, Ferroelectrics, and Frequency Control},
  title     = {Local Sound Speed Estimation for Pulse-Echo Ultrasound in Layered Media},
  year      = {2022},
  issn      = {1525-8955},
  month     = Feb,
  number    = {2},
  pages     = {500--511},
  volume    = {69},
  doi       = {10.1109/tuffc.2021.3124479},
  publisher = {Institute of Electrical and Electronics Engineers (IEEE)},
}

@Article{HeriardDubreuil2023,
  author    = {Hériard-Dubreuil, Baptiste and Besson, Adrien and Wintzenrieth, Frédéric and Cohen-Bacrie, Claude and Thiran, Jean-Philippe},
  journal   = {IEEE Transactions on Ultrasonics, Ferroelectrics, and Frequency Control},
  title     = {Refraction-Based Speed of Sound Estimation in Layered Media: An Angular Approach},
  year      = {2023},
  issn      = {1525-8955},
  month     = Jun,
  number    = {6},
  pages     = {486--497},
  volume    = {70},
  doi       = {10.1109/tuffc.2023.3261541},
  publisher = {Institute of Electrical and Electronics Engineers (IEEE)},
}

@Article{Goicoechea2024,
  author    = {Goïcoechea, Antton and Brütt, Cécile and Le Ber, Arthur and Bureau, Flavien and Lambert, William and Prada, Claire and Derode, Arnaud and Aubry, Alexandre},
  journal   = {Physical Review Letters},
  title     = {Reflection Measurement of the Scattering Mean Free Path at the Onset of Multiple Scattering},
  year      = {2024},
  issn      = {1079-7114},
  month     = Oct,
  number    = {17},
  volume    = {133},
  doi       = {10.1103/physrevlett.133.176301},
  publisher = {American Physical Society (APS)},
}

@Article{Giraudat2024,
  author    = {Giraudat, Elsa and Burtin, Arnaud and Le Ber, Arthur and Fink, Mathias and Komorowski, Jean-Christophe and Aubry, Alexandre},
  journal   = {Communications Earth \&amp; Environment},
  title     = {Matrix imaging as a tool for high-resolution monitoring of deep volcanic plumbing systems with seismic noise},
  year      = {2024},
  issn      = {2662-4435},
  month     = Sep,
  number    = {1},
  volume    = {5},
  doi       = {10.1038/s43247-024-01659-2},
  publisher = {Springer Science and Business Media LLC},
}

@Article{Kawakatsu2009,
  author    = {Kawakatsu, Hitoshi and Kumar, Prakash and Takei, Yasuko and Shinohara, Masanao and Kanazawa, Toshihiko and Araki, Eiichiro and Suyehiro, Kiyoshi},
  journal   = {Science},
  title     = {Seismic Evidence for Sharp Lithosphere-Asthenosphere Boundaries of Oceanic Plates},
  year      = {2009},
  issn      = {1095-9203},
  month     = Apr,
  number    = {5926},
  pages     = {499--502},
  volume    = {324},
  doi       = {10.1126/science.1169499},
  publisher = {American Association for the Advancement of Science (AAAS)},
}

@article{cherkaoui_learning_2020,
	title = {Learning to solve {TV} regularised problems with unrolled algorithms},
	url = {https://dl.acm.org/doi/10.5555/3495724.3496690},
	language = {en},
	number = {966},
	journal = {34th Conference on Neural Information Processing Systems (NeurIPS 2020)},
	author = {Cherkaoui, Hamza and Sulam, Jeremias and Moreau, Thomas},
	month = dec,
	year = {2020},
	pages = {11513 -- 11524},
}

@article{hashimoto_effects_2023,
	title = {Effects of {Size} of {Micro} {Texture} {Regions} on the {Dwell} {Fatigue} {Properties} of {Ti}-{6Al}-{4V}},
	volume = {63},
	issn = {0915-1559, 1347-5460},
	url = {https://www.jstage.jst.go.jp/article/isijinternational/63/10/63_ISIJINT-2023-205/_article},
	doi = {10.2355/isijinternational.ISIJINT-2023-205},
	language = {en},
	number = {10},
	urldate = {2025-09-29},
	journal = {ISIJ International},
	author = {Hashimoto, Shohtaroh and Takebe, Hidenori and Mori, Kenichi and Miyahara, Mitsuo},
	month = oct,
	year = {2023},
	pages = {1774--1785},
}

\end{document}